\pdfoutput=1
\documentclass[12pt]{article}
\usepackage{amsmath}
\usepackage{amsmath}
\usepackage{graphicx}
\usepackage{xr}
\usepackage{lineno}
\usepackage{xcolor}
\usepackage{bm}
\author{Georg~Meisl$^1$, Alexander~J~Dear$^2$, Thomas~C~T~Michaels$^{1,2}$ \\and Tuomas~P~J~Knowles$^{1,3}$\\
	\normalsize{$^1$Centre for Misfolding Diseases, Department of Chemistry,}\\
	\normalsize{University of Cambridge, Lensfield Road, Cambridge CB2 1EW, UK}\\
	\normalsize{$^2$Paulson School of Engineering and Applied Sciences, Harvard University}\\	\normalsize{Cambridge, MA 02138, USA}\\
	\normalsize{$^3$Cavendish Laboratory,  University of Cambridge, 19 JJ Thomson Avenue,}\\	\normalsize{Cambridge CB3 0HE, UK}\\
	\\
}
\date{}

\title{Mechanism, scaling and rates of protein aggregation from in vivo measurements.}
\begin{document}

\maketitle

\noindent\textbf{The formation and proliferation of protein aggregates play a central role in a number of devastating neuro-degenerative diseases. Many experimental studies indicate that the ability of existing aggregates to replicate is a key property in generating their pathogenic effect across a range of diseases. However, given the complexity of the process \textit{in vivo}, no principled general approach currently exists to obtain the rates of the fundamental steps that underlie aggregate formation from measurements in living systems. In order to address this challenge, here we present a general approach for analysing aggregation kinetics that considers broad classes of processes that can be described by a family of scaling solutions. Our approach is not limited only to fibrillar aggregates, but applies to any aggregate shape. 
We show that the rates can reliably be  extracted by fitting of a simple logistic function, even from experimental data in living systems, and give a very general analytical expression that relates the scaling of the exponential rate with monomer concentration to the microscopic details of the underlying reaction. This approach can thus be used to infer the microscopic mechanism driving the aggregation process from macroscopic measurements in complex systems.
}

\section*{Introduction}
Protein aggregation is a phenomenon of broad biological and medical interest, not least because of the central role of such aggregates in many, currently incurable, neurodegenerative diseases\cite{Chiti2006,Knowles2014}. Of particular importance in the context of disease are those aggregates that are able to self-replicate through a multiplication process by which existing aggregates trigger the formation of additional aggregates\cite{Knowles2009,Meisl2020review}. Such aggregates, regardless of their associated disease, are often termed prion-like for this parallel to the highly infectious protein aggregates that are the causative agent of prion diseases\cite{Prusiner1982, Eigen1996,Mudher2017}.

Obtaining detailed mechanistic knowledge of the processes that are responsible for converting normally soluble proteins to their pathological aggregated states is a key objective of many ongoing studies. Significant progress on establishing mechanisms of aggregation of purified protein \textit{in vitro} has been made in the last decades\cite{Knowles2009,Meisl2016}, yet our understanding of how these processes are modulated in living systems is lacking, a factor further complicated by the difficulty of obtaining mechanistic information from\textit{ in vivo} measurements and the fact that even the shape of the growing aggregates \textit{in vivo} is often not clearly established.

\begin{figure}[h]
	\centering
	\includegraphics[width=0.6\columnwidth]{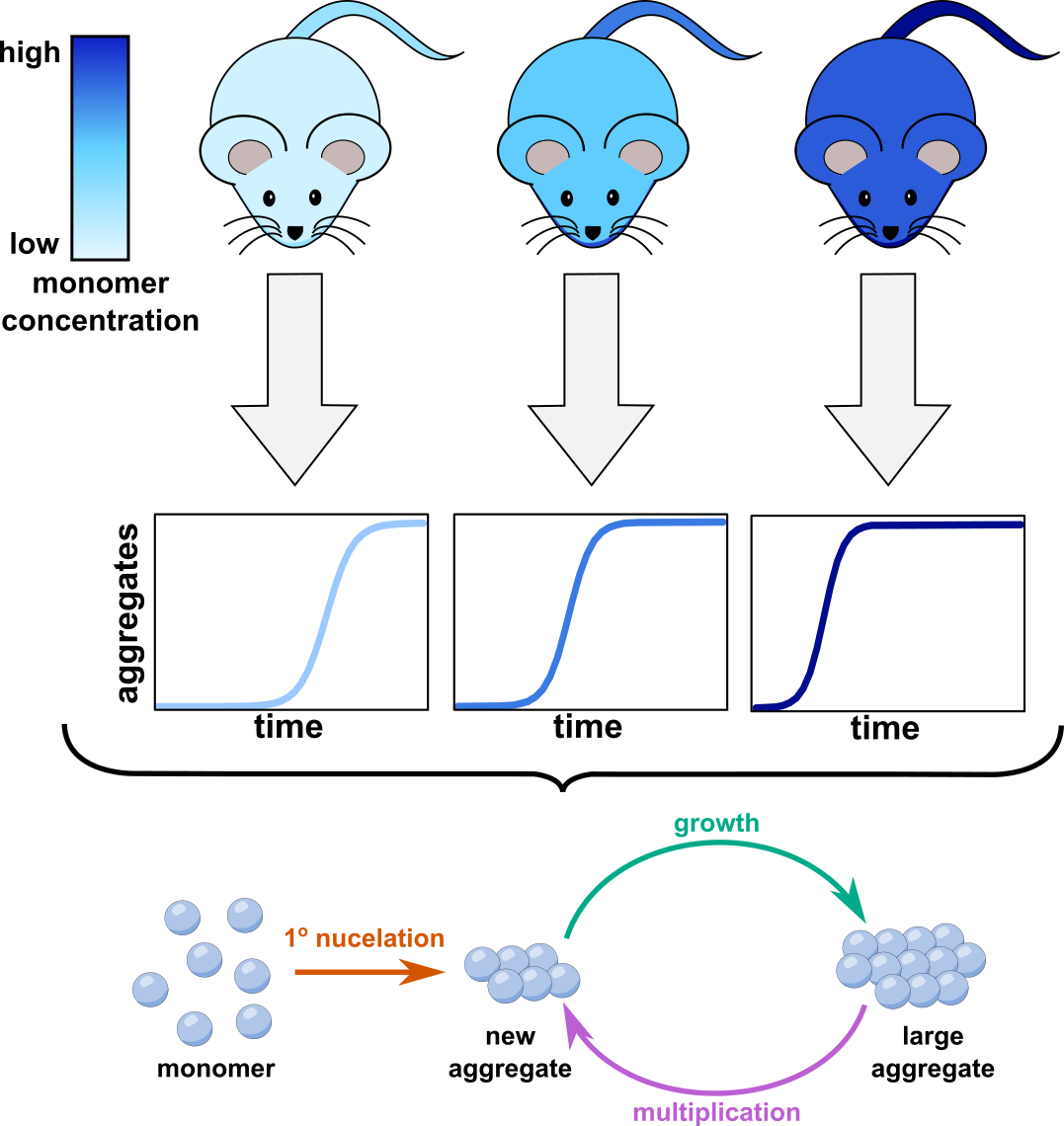}
	\caption{\textbf{Schematic of analysis pipeline.} Measures of aggregate amounts over time in organisms expressing different levels of monomeric protein give rise to aggregation kinetics. Determination of rates of exponential replication and their dependence on the monomer concentration give mechanistic insights.}
	\label{fig:overview}
\end{figure}

In recent decades, a multitude of molecular events have been proposed to govern the aggregation of proteins in living systems\cite{Greer2006, Masel1999b, Kulkarni2003, Come1993, Serio2000, Poeschel2003,  Calvez2009, Payne1998}. A common drawback of many such molecular models is that the level of specificity, in terms of the mechanistic assumptions explicitly and implicitly incorporated in many mathematical models, is often difficult to support given the complex nature of \textit{in vivo} experimental data. The conclusions drawn can thus be very specific and as a result have not gained wide recognition in the research community. In particular, specific assumptions about the aggregate shape are often made and usual aggregates are assumed to be fibrillar and their rates of growth and multiplication are assumed to be size-independent.
 At the other end of the spectrum are purely phenomenological models, which describe the data well and do not suffer from this over-fitting\cite{Jack2010,Jack2013BiomarkerModel,Whittington2018}. However, their fitted parameters typically provide no mechanistic insights into the experimental system, and are merely a means to quantify and interpolate measurements. An approach is therefore needed that is at the same time general enough to not over-interpret noisy and sparse data and still provides a clear molecular interpretation of its parameters\cite{Nowak1998,Eigen1996}. Here we develop such an approach based on our experience of molecular models of aggregation in a multitude of systems. In line with the prion-like nature of many neurodegenerative diseases, we consider mainly those mechanisms that involve the self-replication of aggregates, meaning that the formation of new aggregates happens predominantly by a pathway depending on the presence of existing aggregates. We develop a general approach, that does not require aggregates to be fibrillar and is applicable regardless of the shape of the aggregates, the way in which they grow and the mechanism by which they multiply. We also discuss robust experimentally accessible quantities, the general functional form of the kinetic curves and the scaling of the exponential rate with monomer concentration, that can be used to determine which classes of mechanisms are consistent or not consistent with the data (see Fig.~\ref{fig:overview}).

\section*{Results and Discussion}
The kinetics of the full reaction network underlying an aggregation process is commonly described by a master equation, a set of differential equations that describe how the concentration of aggregates of each size evolves over time\cite{Knowles2009}. We denote by $f(t,j)$ the concentration of aggregates of size $j$ at time $t$ and the general master equation takes the form
\begin{equation}
\frac{df(t,j)}{dt}=\Psi_j[\{f(t,i)\}_i,m(t)] \ \ \ \forall j
\label{eq:master_gen}
\end{equation}
where $\{f(t,i)\}_i$ denotes the set of all $f(t,j)$ and $\Psi_j$ is any functional that may depend on the concentrations of all different sizes of aggregate. The same formalism applies if there are several different species of the same size and in this case $\Psi_j[\cdot]$ is simply a functional of all sizes and types of aggregates.
 The zeroth and first moments of the size distribution are given by $P(t) = \sum_j f(t,j)$ and $M(t) = \sum_jjf(t,j)$. If we treat the monomer concentration separately, i.e. $f(t,1) = m(t)$ and $j\geq2$, then the physical interpretation of $P(t)$ and $M(t)$ is the total number concentration and total mass concentration of aggregates, respectively. These quantities are of particular interest because they, as well as their ratio, the average number of monomers per aggregate $\mu(t)=\frac{M(t)}{P(t)}$, are the most readily accessible experimentally\cite{Meisl2016}. 
Using the master equation, differential equations for these moments, and other quantities that can be expressed as linear combinations of the $f(t,j)$, can be derived and are of the general from
\begin{equation}
\frac{dY}{dt}=F[\{f(t,i)\}_i,m(t)]
\label{eq:dgen}
\end{equation}
where $Y(t)$ is any quantity, such as a $P(t)$, that can be expressed as a linear combination of the $f(t,j)$ and, as above,  $\{f(t,i)\}_i$ denotes the set of all $f(t,j)$ and $F[\{f(t,i)\}_i,m(t)]$ denotes a general functional that may depend on the concentrations of all different sizes of aggregate.

These key observables, $Y$, may depend on the concentration of each aggregated species in an arbitrarily complex manner. This general dependence on the concentration of aggregates of different lengths results in a very difficult general problem. To make progress, we therefore neglect reactions which are second order or higher in the aggregate concentrations. Since the diffusion coefficients and concentrations of the aggregated species are usually smaller than that of the monomer, this is likely to be an accurate assumption and has been found to hold \textit{in vitro}. There are no specific requirements on what the nature of the individual steps is and they can involve any number of monomeric species and their rate can depend on the size of the aggregated species. It is important to note here that this assumption holds for the aggregation of all known amyloid forming proteins \textit{in vitro} and in particular it applies to all common mechanisms of multiplication, such as fragmentation or secondary nucleation\cite{Cohen2013,Meisl2014, Meisl2017a, Gaspar2017}. Interactions between two or more aggregate species may be important for clustering of aggregates at higher concentrations, but such interactions appear to not play a significant role in the kinetics of aggregate formation \textit{in vitro} and are very unlikely to be relevant at the aggregate concentrations encountered \textit{in vivo}, which are several orders of magnitude below those used \textit{in vitro}. Processes that actively sequester aggregates into large plaques or aggresomes \textit{in vivo} are again likely to be of first order with respect to the aggregate concentrations\cite{Park2017,Johnston1998}.   So while physically and biologically this is a relatively conservative assumption, it significantly simplifies the mathematics, as now the rates of change of any species have to be linear in all the $f(t,j)$. In the above equation $\Psi_j[\{f(t,i)\}_i,m(t)]$, and consequently also $F[\{f(t,i)\}_i,m(t)]$, are thus linear functions of the $f(t,j)$, although they are not necessarily linear in $m(t)$.

We now consider the case when the monomer concentration is constant, thus $m(t) = m_0$. This situation is a good approximation for early times also in closed systems where total protein mass is conserved\cite{Knowles2009, Cohen2011a}, and is likely representative of many situations in open biological systems where the monomeric protein concentrations is kept constant by homeostasis\cite{Nowak1998, Masel1999b}. Eq.~\eqref{eq:dgen} now becomes 
\begin{equation}
\frac{dY}{dt}=\sum_{j=2}^{\infty}\alpha_jf(t,j) + \alpha_1
\label{eq:dgen_linear}
\end{equation}
where the $\alpha_i$ may depend on the monomer concentration.

A key observation here is that almost all processes depend on the presence of aggregates, including processes that change the size of existing aggregates by addition or removal of proteins and any processes that change the number of aggregates in a multiplication reaction, such as fragmentation or secondary nucleation. Only the formation of aggregates directly from monomers, primary nucleation does not explicitly depend on the concentration of existing aggregates.

\subsection*{Scaling of replication rate with monomer concentration}

In most cases, the molecular processes that make up the different contributions to the master equation fall exclusively into one of two categories: those that affect only $M(t)$ and those that affect only $P(t)$\cite{Meisl2017a}. Processes that change only the size of existing aggregates without changing their number, i.e. growth or shrinkage, fall into the first category\footnote{shrinkage to a critical size below which the aggregates dissociate into monomers does result in a loss of aggregates, however, as aggregates are generally much larger than these critical sizes, this process is negligible}. By contrast, process by which existing aggregates multiply or new aggregates are formed directly from monomer generally fall into the category of processes that affect only $P(t)$. While some of these processes, such as primary nucleation, may convert some monomers into aggregates, their contribution to $M(t)$ is generally negligible, as the average size of aggregates is much larger than the nucleus sizes. This classification is also evident from the master equation and the definitions of $P(t)$ and $M(t)$.
All processes commonly observed for \textit{in vitro} aggregation can be classified in this manner. However, \textit{in vivo} there may exist processes that completely remove large aggregates from the reaction mixture, through exocytosis, autophagy and other mechanisms. Such processes likely cannot be classified exclusively into affecting only $P(t)$ or $M(t)$.  In the following we will summarize under the term growth the processes that increase $M(t)$ and under the term multiplication the processes that increase $P(t)$, with the exception of primary nucleation, which is considered separately.

Very generally, the rate of growth and its monomer dependence can be different for every aggregate size. Therefore, the general form for the rate of increase in mass $M(t)$ is given by
\begin{equation}
\frac{dM}{dt}=k_g \sum_j g(j,m_0) f(t,j)
\label{eq:dM_general}
\end{equation}
where $g(j,m_0)$ describes the size dependent growth rate (including the reverse reaction, depolymerisation). One interpretation of $g$ is the number of growth competent sites on an aggregate of size $j$, however, this description is also consistent with a growth rate that depends on the size due to other factors. To simplify notation in the following discussions we have also explicitly included a factor $k_g$, which we refer to as the rate constant of growth, so $g$ accounts for the size and monomer dependence of the rate.\footnote{In the standard in vitro model of fibril elongation by monomer addition we would thus have $g(j,m_0)=2m_0$. Also note that in our earlier work\cite{Knowles2009,Cohen2011a,Meisl2017a} we generally use $k_+$ to denote the elongation rate of linear fibrils. We use here $k_g$ to emphasize that this is the rate constant of a general growth reaction, not just linear elongation.} 

 In most cases the same species adds onto all sizes of aggregates and the reaction order of this growth step with respect to the monomer concentration is 1, i.e. $g(j,m_0) = \bar{g}(j)m_0^{n_g}$, with $n_g = 1$. If growth processes are saturated, as can sometimes be seen at particularly high monomer concentrations\cite{Meisl2017a,Dear2020CatNature}, a lower reaction order, $n_g\approx 0$, can be observed. A saturation of the growth process is unlikely in most \textit{in vivo }situations as the concentrations present \textit{in vivo} are well below the concentrations at which saturation of amyloid growth is generally observed in \textit{in vitro} experiments \cite{Collins2004, Buell2010a,Buell2014,Scheibel2004}. Reaction orders of $n_g>1$ could in principle be observed if the species being added is oligomeric, present at low concentrations and in equilibrium with monomer, but the authors are not aware of evidence for such a mechanisms in a real system of protein aggregation.

In analogy to Eq.~\eqref{eq:dM_general} above, the general form for the increase in the number of aggregates is given by 
\begin{equation}
\frac{dP}{dt}=k_m \sum_{j}\sigma(j,m_0)f(t,j) 
\label{eq:dP_general}
\end{equation}
These multiplication processes are also often classified into two categories, those whose rate does depend on monomer concentration and those whose rate does not depend on monomer concentration. For example, multiplication by secondary nucleation generally yields higher reaction orders with respect to the monomer concentration\cite{Cohen2011a, Cohen2013,Ferrone1985b}, whereas fragmentation is usually independent of the monomer concentration. \footnote{Note that in our earlier work\cite{Knowles2009,Cohen2011a,Meisl2017a} we generally use $k_2$ to denote the rate constant of secondary nucleation and $k_-$ to denote the rate constant of fragmentation. They are here replaced by $k_m$ to emphasize that we consider more general multiplication reactions.} 
Normally one type of process dominates the aggregation reaction over a range of monomer concentrations, but in rare cases a competition of two multiplication processes can be observed\cite{Meisl2017a}. Any such phenomenon of multiple processes competing can easily be included in the $\sigma(j,m_0)$ in the above description.
In addition to these multiplication processes, a primary nucleation process can be included in the form of a term that depends only on the monomer concentration and is independent of the $f(t,j)$. However, under conditions of constant monomer, primary nucleation is simply a source term and in the presence of multiplication processes it will always become negligible as the late time limit is approached, because the contribution of multiplication to the rate increases with increasing aggregate concentration in a positive feedback loop.

The aim now is to use the general equations for $dP(t)/dt$ and $dM(t)/dt$ above to determine an approximate general solution and its dependence on the concentration of monomer, $m_0$. We rewrite $f(t,j) = P(t)\rho(t,j)$, where $\rho(t,j)$ now denotes the normalised size distribution to get
\begin{eqnarray}
\frac{dM}{dt}&=&k_g P(t) \sum_j g(j,m_0) \rho(t,j)\nonumber \\
\frac{dP}{dt}&=&k_mP(t) \sum_{j}\sigma(j,m_0)\rho(t,j)
\end{eqnarray}
As $\rho$ is now a normalised distribution, the sums in these expressions are expectation values of the functions $g$ and $\sigma$. In order to obtain a closed from expression, we approximate the expectation value of a function by the function of the expectation value of its argument, i.e. $\langle g(j) \rangle \approx g(\langle j \rangle)$. This approximation will be most accurate when the size distribution has most of its probability mass close to its mean and when the $g$ is close to linear in $j$. The performance of this approximation is evaluated in a number of different systems below. The expectation value $\langle j \rangle$ is simply the average length so can be replaced by $\mu(t)=\frac{M(t)}{P(t)}$.

\begin{eqnarray}
\frac{dM}{dt}&=&k_g P(t) g(\mu(t),m_0)\label{eq:dM_pre_ss} \\
\frac{dP}{dt}&=&k_mP(t) \sigma(\mu(t),m_0) \label{eq:dP_pre_ss}
\end{eqnarray}
Now in order to proceed and derive a general solution, we would like to remove the time-dependence of the average length $\mu(t)$. In other words, we would like to consider a steady state size-distribution (overall aggregate amounts may still vary). Such a steady state size-distribution arises when, after an initial adjustment period, a single solution mode dominates the system of ODEs that describes the size distribution kinetics. In turn, this occurs when at least one mode undergoes exponential growth, leading to an exponentially increasing aggregate mass. Indeed many systems show this exponential increase under constant monomer and quickly approach such a time-invariant size distribution after a period determined by initial conditions\cite{Michaels2015,Cohen2011a}. It thus remains to establish the general conditions under which exponential growth of $f(t,j)$ emerges. We show in the Appendix that a necessary condition for exponential increase is that there exists at least one feedback term in the size distribution equations, such that larger aggregates generate smaller ones, for example through fragmentation or secondary nucleation. Given the evidence that aggregates with and ability to self-replicate are the key aggregate species in most aggregation-related disorders\cite{Mudher2017, Meisl2020arxivprions, Purro2018, Iba2013}, a multiplication step will likely be present for most systems of interest.

Therefore, a steady state size distribution is likely to be approached by a wide range of systems, and under this assumption of steady state, the average length remains constant, $\mu(t) = \mu$, and the moment equations can be solved to give
\begin{eqnarray}
P(t)&=&P_0e^{(k_m \sigma(\mu,m_0))t}\nonumber \\
M(t)&=&\frac{k_g g(\mu,m_0)}{k_m \sigma(\mu,m_0) } P_0e^{(k_m \sigma(\mu,m_0))t} 
\end{eqnarray}
where $P_0 = P(0)$ and the exponential rate is
\begin{equation}
\kappa =k_m \sigma(\mu,m_0)
\label{eq:kappa_gen}
\end{equation}
In order for this to be self-consistent, we require $\mu=\frac{M(t)}{P(t)}$, which allows us to determine $\mu$.
\begin{equation}
\mu=\frac{k_g g(\mu,m_0)}{k_m \sigma(\mu,m_0)}
\label{eq:mu_gen}
\end{equation}
We note that the numerator represents the rate of growth, whereas the denominator represents the rate of multiplication. Therefore, as $\kappa$ depends on both rates, experimental measurements of $\mu$ are needed to obtain estimates of the relative rates of growth and multiplication.
The scaling of the rate $\kappa$ is defined as
\begin{equation}
\gamma=\frac{d}{ \log(m_0)}(\log(\kappa))=m_0\frac{d \log(\sigma(\mu,m_0)}{ dm_0}
\label{eq:scaling_gen}
\end{equation}
so if the specific functions $g$ and $\sigma$ are known, $\mu$ can be calculated and thus the scaling can be determined.

These expressions are still general and also apply in the presence of processes that remove aggregates from the reaction. The key questions in the context of clearance mechanisms, however, concern when they are sufficient to prevent run-away aggregation. In the approach we present here, we consider systems in which the aggregate amounts are exponentially increasing, and thus clearance is necessarily a less important process. We therefore do not explicitly include clearance processes in the following discussion and instead will present a detailed discussion of their effect in a separate work.
	
To summarise, under the basic assumptions that aggregation does not involve reactions of more than one aggregate and the aggregate mass grows exponentially,  simple closed form expressions for the exponential rate and average aggregate size can be derived, Eqs.~\eqref{eq:kappa_gen} and \eqref{eq:mu_gen}. 
In the rare cases when the increase in mass is experimentally observed to not be initially exponential, the kinetics of these systems may be governed by processes other than growth-multiplication, such as the direct formation of aggregates from monomer by primary nucleation, or effects of spatial transport (see later) may be rate-limiting.

In the following, we will illustrate our approach by considering generalised descriptions of growth and fragmentation, derive analytical solutions in the limiting cases, and use numerical integration of the master equations in the remaining regions to determine the scaling and show that the values we obtain are consistent with the approximate general expressions obtained in Eqs.~\eqref{eq:kappa_gen} and \eqref{eq:mu_gen}.

\subsection*{Specific parametrisation of size dependences of growth and fragmentation}

To illustrate how the moment equations described previously arise from the master equation, we consider growth by monomer addition with a  generalised size dependence of the rate, along with a generalised fragmentation process.
\begin{eqnarray}
\frac{df(t,j)}{dt}&=&k_g m_0 \left(g(j-1)f(t,j-1)-g(j)f(t,j)\right) \nonumber \\
&&+2\sum_{i=j}^{\infty}k(i,j)f(t,i)- f(t,j)\sum_{i=1}^jk(j,i) \label{eq:full_dist}
\end{eqnarray}
where, as before, $m_0$ is the monomer concentration, $k_g$ is the rate constants of growth, $g(j)$ describes the size-dependence of growth and $k(i,j)$ is the fragmentation kernel. The fragmentation term has been expressed in its most general form, where the fragmentation of an aggregate of size $i$ into an aggregate of size $j$ is given by $k(i,j)$, a rate constant which in general depends both on $i$ and $j$ (the first argument being the size of the original fibril, the second argument the size of one of the fragments.\footnote{As fragmentation yields a piece of size $j$ and another one of size $i-j$, we require $k(i,j)=k(i,i-j)$. This is reflected by the factor of 2 in the second line; an aggregate of size $i$ can create an aggregate of size $j$ either by breaking off a piece of size $j$ or one of size $i-j$.}) We assume that the fragmentation into pieces that are too small to grow and dissociate instead is negligible.
The zeroeth and first moments of this distribution give the number concentration, $P$, and mass concentration, $M$,  of aggregated material, thus
\begin{eqnarray}
\frac{dP(t)}{dt}&=&\sum_{j=1}^{\infty}{\frac{df(t,j)}{dt}}\approx k_m\sum_{j=1}^{\infty}\phi(j)f(t,j) \label{eq:dN_gen_SI}\\
\frac{dM(t)}{dt}&=&\sum_{j=1}^{\infty}{j\frac{df(t,j)}{dt}}=k_g m_0 \sum_{j=1}^{\infty}g(j)f(t,j)
\label{eq:dM_gen_SI}
\end{eqnarray}
where in Eq.~\eqref{eq:dN_gen_SI} the elongation term cancels (as expected because elongation does not affect the number of aggregates) and in Eq.~\eqref{eq:dM_gen_SI} the fragmentation term cancels (as expected because fragmentation does not change the amount of aggregated material, ignoring loss back to monomer due to fragmentation). The fragmentation term has been simplified by defining $\sum_{i=1}^jk(j,i)=k_m\phi(j)$. Note that although the moment equation depends explicitly only on the sum $\sum_{i=1}^jk(j,i)$, which is the rate at which an aggregate of size $j$ fragments, the size distribution $f(t,j)$ will depend on the full fragmentation kernel, $k(j,i)$.

The most notable difference to the case of linear polymerisation\cite{Knowles2009, Cohen2011a} is that these moment equations, \eqref{eq:dN_gen_SI} and \eqref{eq:dM_gen_SI}, are not closed and still depend on the aggregate size distribution $f(t,j)$, making them very difficult to solve. In the special cases that $g(j) =j$ or $g(j)=1$ and $\phi(j) =j$ or $\phi(j)=1$, the above equations are again closed and straightforward to solve.

\subsubsection*{Aggregate growth}
The rate of growth can, in general, depend on the size of the aggregate, which we accounted for with the factors $g(j,m_0)$ above. While other factors may influence the size-dependence of the rate of growth, we explore here specifically the change in the number of growth-competent sites per aggregate due to its dimensionality. If all sizes grow by addition of the same type of species, we can make the implicit dependence of $g$ on the monomer concentration explicit and replace $g(j,m_0) \rightarrow m_0^{n_g}\bar{g}(j)$. For the following example we assume the species that adds is monomeric and adds in an unsaturated reaction, so $n_g=1$. In the case of linear aggregates, the number of growth competent sites is independent of aggregate size, thus $\bar{g}(j) = 2$ (for 2 growing ends) and $\sum_j \bar{g}(j) f(t,j)=2P(t)$ as encountered for linear polymerisation \cite{Knowles2009, Cohen2011a}.

\begin{figure}[h]
	\centering
	\includegraphics[width=0.4\columnwidth]{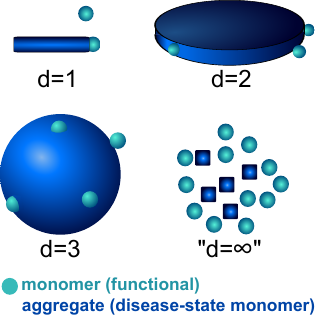}
	\caption{\textbf{Illustration of specific aggregate geometries.} Our formalism can account for a size-dependent growth rate through the parameter $d$. One property that gives rise to a size-dependent growth rate is the dimensionality of the growing aggregate; linear aggregates grow independent of the size ($d=1$), whereas higher dimensional aggregates, such as discs ($d=2$) or spheres ($d=3$) show a size dependent growth rate. The mathematical limit of $d\rightarrow \infty$ corresponds to a direct conversion of functional monomer to a disease state monomer, without aggregation.}
	\label{fig:dimensionality_comic}
\end{figure}

However, as the dimensionality of an aggregate increases, for example from linear to planar, the number of growth competent sites per aggregate will depend on the size of the aggregate. One dimensional aggregates, i.e. fibrils, grow from the ends and the number of growth competent sites per aggregate does not change with size. Two dimensional aggregates by contrast can grow at any point along their edge hence the number of growth competent sites scales as the circumference of the aggregate, (which is proportional to number of monomers in the aggregate to the power of $1/2$). To describe the generalised behaviour here we introduce the dimensionality $d$, which is a number $\geq 1$. This formulation allows us to derive a general model, which is a continuous function of the dimensionality. Intermediate values of $d$, e.g. 1.2 can account for slightly branched linear aggregates. More generally, $d$ can simply be considered as a parameter that encapsulates the size dependence of the growth rate, due to any effect in which the rate depends on a power of the size. An illustration of some possible geometries of growing aggregates and their corresponding values of $d$ is shown in Fig.~\ref{fig:dimensionality_comic}.

The number of growth-competent sites of and aggregate of size $j$ is related to its mass as follows
\begin{equation}
\bar{g}(j)=a_dj^{(1-1/d)}
\label{eq:growth_sites}
\end{equation}
where $d$ is the dimensionality and $a_d$ is a shape-dependent constant of proportionality. The values of $a_d$ in some special cases are
\begin{eqnarray}
&=&2\mathrm{\ for\ linear \ aggregates} \nonumber \\
a_d&=&2 \sqrt{\pi}\mathrm{\ for\ disc\ shaped\ aggregates} \nonumber \\
&=&3^{2/3}(4\pi)^{1/3}\mathrm{\ for\ spherical\ aggregates} \nonumber
\end{eqnarray}
In the following we will subsume $a_d$ into the elongation rate constant $k_g$. 

The limits of interest are $d=1$ which gives $\sum_j \bar{g}(j) f(t,j)=P(t)$ and $d\rightarrow \infty$ which gives $\sum_j \bar{g}(j) f(t,j) = M(t)$.
Physically, a value of $d=1$ corresponds to the linear polymerisation regime, i.e. each aggregate adds mass to itself at a constant rate, independent of its size. This is the typical behaviour for amyloid fibrils, such as A$\beta$, \textit{in vitro}. Regarding intermediate values of $d$, a spherical aggregate would be described by $d=3$, but structures of a 3 dimensional, highly branched nature could, up to a certain size, result in $d>3$. The limit of $d\rightarrow\infty$ represents a hypothetical case in which the elongation rate of a single aggregate is proportional to its mass. Every monomer in the aggregate converts more monomer into aggregate. In this limit any fragmentation processes become irrelevant, in fact the very concept of separate aggregates is meaningless.  Physically this model describes the kinetics of a prion-like disease if it were merely due to individual proteins changing their conformational state to a ``prion-state". Each one of these ``prion-state" proteins can convert more native proteins to their prion state in this exponential run away reaction. This is the hetero-dimer mechanism originally proposed by Prusiner et al \cite{Prusiner1982} to describe prion replication. It is crucial to point out that, physically, this is significantly different to a description involving aggregation: If aggregates are formed, only a fraction of the aggregated ("prion-state") proteins, specifically the growth competent, accessible sites, can convert more soluble protein to its prion state.

\subsubsection*{Fragmentation}

As fragmentation is independent of the monomer concentration, we replace $\sigma(j,m_0) \rightarrow \bar{\sigma}(j)$. Eq.~\eqref{eq:dN_gen_SI} depends on the entire size distribution, not just its moments, a fact which originates from the general treatment of the fragmentation rate. The form of this fragmentation kernel is in general unknown, so in order to advance with this description an explicit assumption about the functional form of $k(i,j)$ has to be made. Following previous work \cite{Krapivsky_book}, we assume that the probability of an aggregate of size $j$ to fragment, $k_m\bar{\sigma}(j)=\sum_{i=1}^jk(j,i)$, is proportional to the aggregate size to some power $\alpha$, $\bar{\sigma}(j)=j^{\alpha}$, where $\alpha$ is referred to as the homogeneity index. Note that this merely constitutes an assumption about the functional form of the fragmentation kernel and no specific assumptions about the individual $k(j,i)$ need to be made.

The exponent $\alpha$ describes the dependence of the fragmentation rate on the size of the aggregate. The limits of interest are $\alpha=1$ which gives $\sum_j \bar{\sigma}(j) f(t,j)=M(t)$ and $\alpha=0$ which gives $\sum_j \bar{\sigma}(j) f(t,j) = P(t)$. A value of $\alpha=1$ means the probability of an aggregate to fragment is proportional to its size. Such a description emerges in the case of linear polymerisation, assuming that an aggregate is equally likely to break at any monomer-monomer connection. This scaling is expected to decrease as the dimensionality of the aggregate increases (for non-linear aggregates, larger aggregates require more bonds to be broken to fragment). In the limit of $\alpha=0$ each aggregate is equally likely to fragment, regardless of its size. A value of $\alpha$ below 0 corresponds to what is known as the shattering region\cite{Krapivsky_book}, where small aggregates fragment explosively fast, and was therefore not considered in this description. 

Situations in which $\alpha>1$, i.e. where the probability per monomer to break increases with aggregate size, could be envisioned when the fragmentation effects act on larger scales than single bonds, for example when mechanical stress affects aggregates above a certain size, such as aggregation under shaking.

Putting together the functional forms for $g(j) = j^{1-1/d}$ and $\phi(j)=j^{\alpha}$ we obtain the moment equations
\begin{eqnarray}
\frac{dP(t)}{dt}&=&k_m\sum_{j=1}^{\infty}j^{\alpha}f(t,j) \label{eq:dP_frag_growth}\\
\frac{dM(t)}{dt}&=&k_g m_0 \sum_{j=1}^{\infty}j^{1-1/d}f(t,j)
\label{eq:dM_frag_growth}
\end{eqnarray}
which we will now solve for some limiting cases.

\subsubsection*{Solutions for limiting cases}
In the case that $\alpha$ is chosen such that $\sum_jf(t,j)k_mj^{\alpha}$ is a moment of the distribution (e.g. $\alpha=0$ or $\alpha=1$) the explicit dependence of the moment equations on the individual $k(j,i)$ disappears.
Similarly, Eq.~\eqref{eq:dM_gen_SI} becomes an equation that depends only on the moments, not the entire size distribution, in the case that $d=1$ and the limit that $d\rightarrow\infty$. 
Therefore, there are 4 possible combinations of limiting cases, that yield a closed, solvable system of moment equations, which will be discussed below (in the limit of $d\rightarrow\infty$, any multiplication processes become negligible, hence the cases of $\alpha=0$ and $\alpha=1$ are equivalent for $d\rightarrow\infty$). The approximate scaling predicted by Eqs.~\eqref{eq:mu_gen} and \eqref{eq:scaling_gen}. As $g(j,m_0) = m_0j^{1-1/d}$ and $\sigma(j,m_0)=j^{\alpha}$ we obtain 
\begin{equation}
\mu=\frac{k_g m_0\mu^{1-1/d}}{ k_m \mu^{\alpha}} \ \ \ \ \mathrm{and\ thus}\ \ \ \ \mu=\left(\frac{k_g m_0}{ k_m}\right)^{\frac{1}{\alpha+1/d}}
\label{eq:mu_elon_frag}
\end{equation}
and thus
\begin{equation}
\kappa=k_m\left(\frac{k_g m_0}{ k_m}\right)^{\frac{\alpha}{\alpha+1/d}}
\end{equation}
Therefore the scaling is given by
\begin{equation}
\gamma=\frac{\alpha}{\alpha +1/d}
\label{eq:scaling_elon_frag}
\end{equation}
In the limiting cases this gives $\gamma =1$ as $d\rightarrow \infty$, $\gamma =1/2$ for $\alpha = d=1$ and $\gamma =0$ for $\alpha=0$ and $d=1$.

\noindent\textbf{Linear growth and breakage ($d=1,\ \alpha=1$):} This is the case of linear polymerisation with fragmentation proportional to the mass, which has been solved previously for a closed system with monomer addition\cite{Cohen2013, Meisl2016}. 
The moment equations are
\begin{eqnarray}
\frac{dP(t)}{dt}&=&k_m M(t) \\
\frac{dM(t)}{dt}&=&2k_g m_0 P(t)
\end{eqnarray}
which admit as a solution
\begin{eqnarray}
M(t)&=&M_0\cosh(\kappa t) + P_0\sqrt{\frac{2k_g m_0}{k_m}}\sinh(\kappa t)\nonumber \\
P(t)&=&P_0\cosh(\kappa t) + M_0\sqrt{\frac{k_m}{2k_g m_0}}\sinh(\kappa t)\nonumber
\end{eqnarray}
where the exponential rate parameter $\kappa = \sqrt{k_gm_0k_m}$, is the same as that obtained from the expression in Eq.~\eqref{eq:kappa_gen}. In the long time limit the scaling is thus $\gamma=1/2$ and the average length $\mu = \frac{M(t)}{P(t)}=\sqrt{\frac{2k_g m_0}{k_m}}$, also consistent with the expression derived in Eq.~\eqref{eq:mu_gen}.

\noindent\textbf{Size-independent multiplication ($d=1,\ \alpha=0$):} This is the case of linear polymerisation where each aggregate is equally likely to fragment, regardless of its size. The moment equations are given by
\begin{eqnarray}
\frac{dP(t)}{dt}&=&k_mP(t)\\
\frac{dM(t)}{dt}&=&k_g m_0 P(t)
\end{eqnarray}
which are solved by
\begin{eqnarray}
P(t)&=&P_0e^{k_mt}\nonumber \\
M(t)&=&\frac{k_g m_0}{k_m}P_0e^{k_mt}\nonumber
\end{eqnarray}
where $\kappa=k_mf$, $\mu = \frac{k_g m_0}{k_m}$ and $\gamma=0$, all consistent with our general expressions.

\noindent\textbf{High dimensional growth ($d\rightarrow\infty$):}
This is the limit in which every converted monomer converts more monomer, thus we also refer to this mechanism as direct conformational conversion (such a  mechanism was originally proposed by Prusiner as a model for prion replication\cite{Prusiner1982}). The distinction of separate aggregates becomes pointless and hence so does a distinct treatment of nucleation or fragmentation (thus $M(t)=P(t)$ and$k_gm_0 = k_m$). There is only one moment equation:
\begin{eqnarray}
\frac{dP(t)}{dt}&=&k_mP(t)
\end{eqnarray}
which is solved by
\begin{eqnarray}
P(t)&=&P_0e^{k_mt}
\end{eqnarray}
giving a exponential rate of $\kappa=k_m$, an average length of $\mu = 1$ (because there is no aggregation so $P(t)=M(t)$) and $\gamma=1$, all consistent with our general expressions.

In summary an increase in $d$ leads to an increase in the contribution of elongation to the scaling exponent, whereas a decrease in $\alpha$ leads to decrease of the contribution of elongation to the scaling exponent, by decoupling fragmentation from elongation.

\subsection*{Numerical interpolation between limits}
\label{sec_numerics_SI}
In order to estimate how the scaling exponent varies for  values of $d$ and $\alpha$ which do not have an analytical solution, we numerically integrated the master equation describing the fibril length distribution, Eq.~\eqref{eq:full_dist}. In this context a fourth order Runge-Kutta integration algorithm was employed, the time evolution of the fibril length distribution was determined at 40 different $d$ and $\alpha$ values, at 4 different monomer concentrations each. In order to evaluate numerically Eq.~\eqref{eq:full_dist}, the full fragmentation Kernel $k(j,i)$ needs to be specified. Given that $\sum_{i=1}^jk(j,i)=k_m\sigma(j)=k_mj^{\alpha}$ is fixed by our definition of the fragmentation homogeneity, we let $k(j,i)=k_mj^{\alpha}/j$, i.e. the production of a piece of any size is of equal probability.  As the master equation consists of infinitely many differential equations, one equation for each fibril length, a cut-off has to be chosen in order to make numerical integration possible. In this case it was chosen at a fibril length of 50000. The average fibril length in all simulations remained well below that value, suggesting that the cut-off does not influence the result of the integration.
\begin{figure}[h]
	\centering
	\includegraphics[width=0.7\columnwidth]{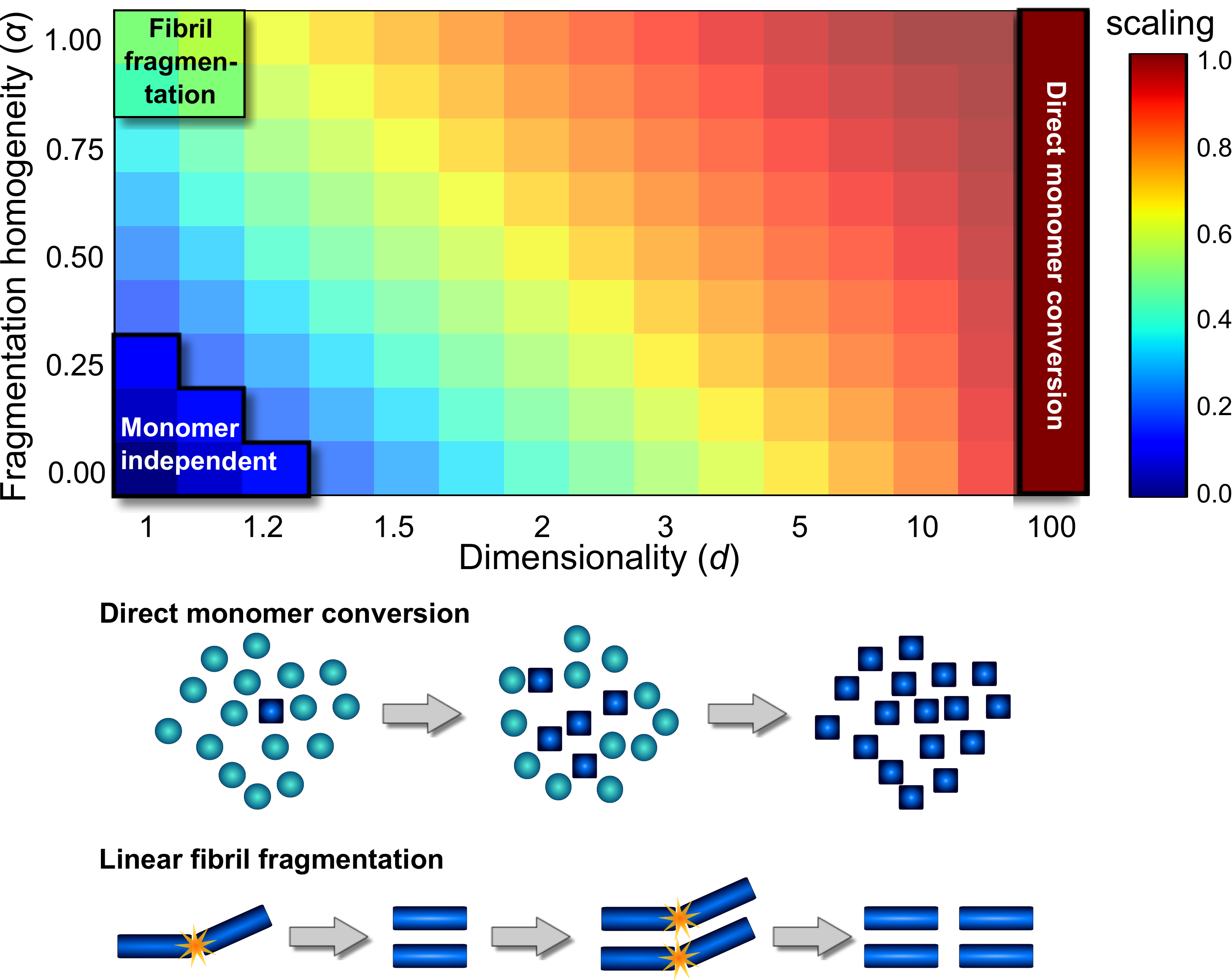}
	\caption{\textbf{Scaling exponent in different limits.} The scaling exponent, evaluated by numerical integration of the moment Eqs.~\eqref{eq:dP_frag_growth} and \eqref{eq:dM_frag_growth} for generalised growth and fragmentation, is shown for a number of different size dependences of growth and fragmentation. The different limiting regimes are highlighted and schemes for the corresponding mechanisms shown below. The approximate expression derived for the scaling, from Eqs.~\eqref{eq:kappa_gen}, \eqref{eq:mu_gen} and \eqref{eq:scaling_gen}, yields a highly similar plot.  }
	\label{fig:heatmap}
\end{figure}

The full length distribution was then used to calculate the time evolution of fibril number and fibril mass. These results were treated in the same manner as experimental data would: an exponential was fitted to the time evolution of aggregate mass in order to extract $\kappa$. The monomer dependence of the 4 values of $\kappa$ at each $d$ and $\alpha$ combination was then used to determine the scaling $\gamma$. The resulting values are plotted in Fig.~\ref{fig:heatmap}, which was smoothed by linear interpolation between adjacent points. The values chosen for the parameters were $k_m=2\cdot10^{-5}$ s$^{-1}$ and $k_g=10^8$ s$^{-1}$M$^{-1}$, 
 the kinetics were evaluated at 4 different monomer concentrations, spanning an order of magnitude around 1nM. To within error, the simulations agree with our general approximate scaling calculated above, Eqs.~\eqref{eq:kappa_gen}, \eqref{eq:mu_gen} and \eqref{eq:scaling_gen}.

\subsection*{Factors limiting aggregation and late-time plateau behaviour}
In any physical situation, exponential growth is only possible for a finite amount of time. Indeed, the appearance of a plateau value is very common in all experimental systems, and curves of aggregate concentrations over time are usually sigmoidal. The origin of this plateau is easily explained by conservation of total protein mass in a closed system, such as the aggregation of purified protein in a test tube\cite{Knowles2009, Cohen2011a}, but in more complex systems and \textit{in vivo} plateaus may also arise for a variety of other reasons\cite{Payne1998,Jack2013AbPlateau}. The aim of this work is to determine the replication rate and mechanistic class of the aggregation reaction, thus the initial exponential increase is most informative and the specifics of the approach to the plateau are of little importance. In the context of the related diseases, this exponentially increasing phase is also the most relevant as it is responsible for the increase in the number of aggregated species by orders of magnitude.  We therefore propose to use a logistic function to fit experimental data in practice, Eq.~\eqref{eq:logistic}, as it recovers the initial exponential behaviour and allows for plateauing at late time. 
\begin{equation}
M(t)=M_{\infty}\frac{1}{1+e^{-\kappa (t-t_h)}}
\label{eq:logistic}
\end{equation}
where $M_{\infty}$ is the plateau value, $\kappa$ is the replication rate as in Eq.~\eqref{eq:kappa_gen} and $t_h$ is the half time, which contains information about the initial concentration of aggregates and the primary nucleation rate, but which we do not explicitly link to the underlying microscopic processes here.

\subsection*{Performance in experimental systems}
To showcase the application of this approach we first use a well-controlled system obtained by simulating data of the \textit{in vitro} aggregation of A$\beta$42, and then discuss performance for experimental data from a significantly more complex \textit{in vivo} system below. 
These artificial data were generated by numerical integration of the moment equations,
\begin{eqnarray}
\frac{dP(t)}{dt}&=&k_nm(t)^{n_c}+k_2m(t)^{n_2}M(t)\\
\frac{dM(t)}{dt}&=&2k_+ m_(t) P(t)
\end{eqnarray}
under conservation of mass ($m(t) +M(t)=m_0$), which describe the formation of fibrillar aggregates through a double nucleation mechanism. Primary nucleation generates new aggregates directly from monomer with a reaction order of $n_c=2$, secondary nucleation generates new aggregates from monomers, on the surface of existing aggregates with reaction order $n_2=2$ and elongation proceeds by the addition of monomers to the ends of fibrils as described by a single-step mechanism. This model and the rates constants used are those obtained by Cohen et al.\cite{Cohen2013} from the analysis of the \textit{in vitro} aggregation of purified A$\beta$42 peptide, which is linked to Alzheimer's disease.
\begin{figure}[h!]
	\centering
	\includegraphics[width=\columnwidth]{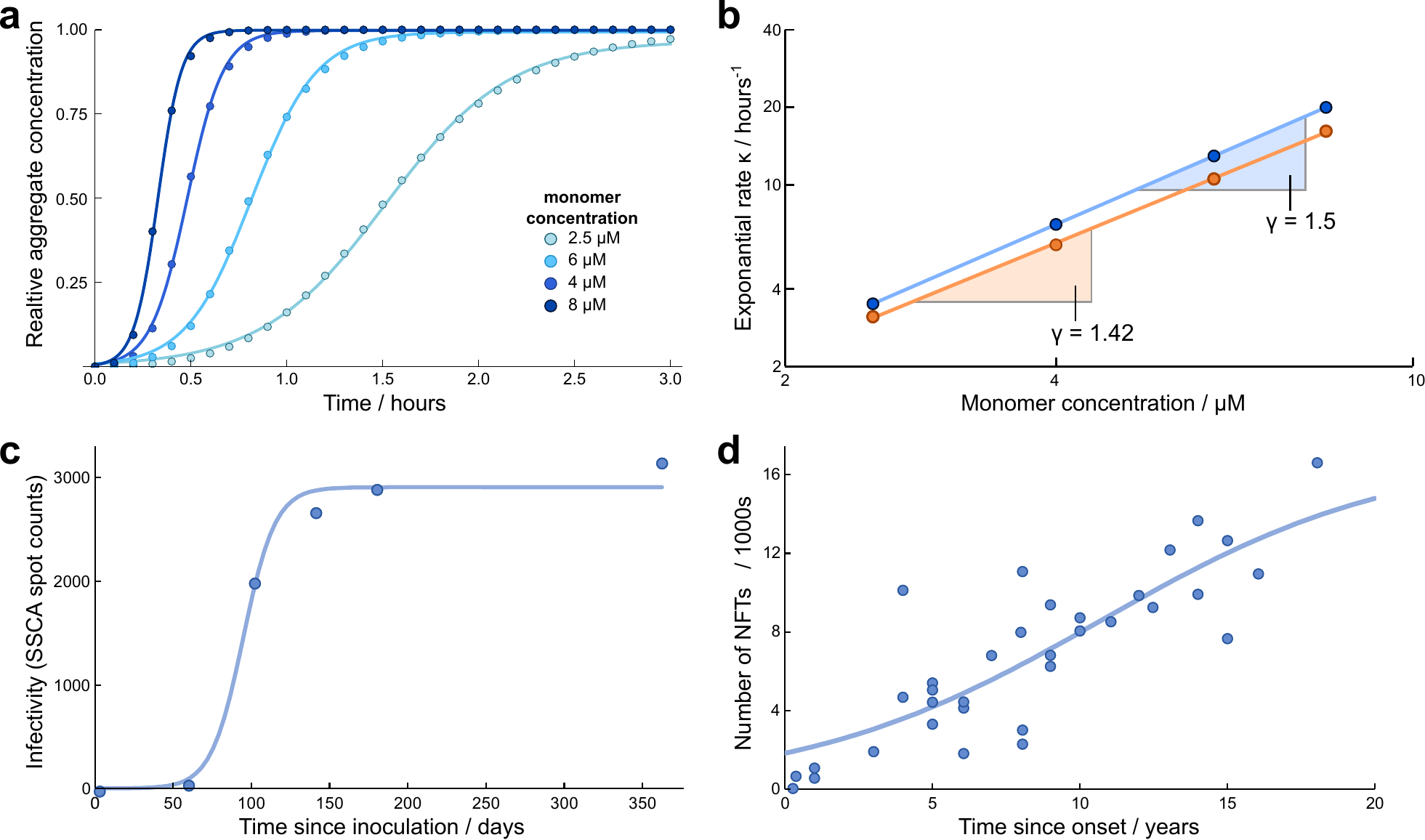}
	\caption{\textbf{Logistic fit to solution of moment equations.} \textbf{a} The data to be fitted (points) were generated by numerical integration of the moment equations for aggregation with primary and secondary nucleation, with parameters set to  $k_n=3\cdot 10^{-4}$ M$^{-1}$s$^{-1}$, $k_+=3\cdot 10^{6}$ M$^{-1}$s$^{-1}$, $k_2=10^{4}$  M$^{-2}$s$^{-1}$ and $n_c=n_2=2$, at a range of initial monomer concentrations. Logistic functions, Eq.~\eqref{eq:logistic}, were then fitted to each kinetic curve individually. \textbf{b} The exponential rates obtained from the fits in (a) (red points) compared to the prediction of the exponential rate from Eq.~\eqref{eq:kappa_gen}, using the set values of the rate constants $k_+$ and $k_2$ and the reaction order $n_2$. \textbf{c} Fits of Eq.~\eqref{eq:logistic} to the infectivity in prion infected mice measured by the Standard Scrapie Cell Assay from  Mays \textit{et al.} \cite{Mays2015}. \textbf{d} Fits of Eq.~\eqref{eq:logistic} to post-mortem stereological counts of the number of neurofibrillary tangles in the superior temporal sulcus of Alzheimer's disease patients from  Gomez-Isla \textit{et al.}\cite{Gomez-Isla1997}.  }
	\label{fig:logistic_fit}
\end{figure}
Note that the above moment equations contain a primary nucleation term, which we have not included in our approach as it becomes negligible at late times. Therefore, our approximations break down when primary nucleation becomes the dominant generator of new aggregates, i.e. when the majority of new aggregates is generated by primary nucleation, not multiplication. Given the rate constants used here, primary nucleation produces more aggregates than secondary nucleation as long as $M(t) \leq 0.03$ $\mu$M $=M_{\mathrm{crit}}$. Thus it always dominates the initial trajectory for reactions starting from a  purely monomeric sample, but how quickly $M_{\mathrm{crit}}$ is reached, i.e. when secondary nucleation takes over depends on the monomer concentration. At lower monomer concentrations, primary nucleation dominates for a larger portion of the time until completion of the reaction. Indeed, as can be seen in Fig.~\ref{fig:logistic_fit}a, the fits of the logistic equation are poorest at the lowest concentration. Eq.~\eqref{eq:kappa_gen} predicts an exponential rate of $\kappa = \sqrt{2k_+k_2m_0^{n_2+1}}$ and a scaling of $\gamma = 1.5$, our fits, Fig.~\ref{fig:logistic_fit}, yield $\gamma = 1.42$ and, evaluated at 4 $\mu$M $\kappa =6$ s$^{-1}$ compared to $\kappa =7$ s$^{-1}$ from Eq.~\eqref{eq:kappa_gen}. Thus, the use of a logistic function is sufficient to account for the late time plateau, in this case due to monomer depletion, and the fits match the aggregation curves well even when there are additional primary nucleation processes. The rates and scaling extracted in this way is consistent with the interpretation of these quantities in terms of the underlying microscopic rate constants given in Eq.~\eqref{eq:kappa_gen}.

Equally, the same strategy can be applied to experimental data obtained in a complex \textit{in vivo} system, as we demonstrate using data of prion disease in mice in Fig.~\ref{fig:logistic_fit}c and Alzheimer's disease in humans in Fig.~\ref{fig:logistic_fit}c. The prion data were obtained from Mays \textit{et al.} \cite{Mays2015} and show measurements of the concentration of the infectivity in heterozygous Prnp$^{+/-}$ mice. In this system, mice expressing the PrP protein are inoculated intra-cerebrally with a sample containing prions, infectious particles consisting mainly of an aggregated from of PrP, which then trigger the aggregation of more PrP in the mouse. Mice are taken at several timepoints after inoculation and measurements are performed by homogenising their brains and determining its infectivity in a cell based assay. The infectivity readout is a measure for the concentration of infectious units. We assume that they constitute a subset of the aggregated species and therefore, under the assumption of a steady state distribution, the time evolution of infectivity can be approximated by the same functional form as that of the total aggregate concentration. We obtain $\kappa=0.1$ days$^{-1}$, which corresponds to the number of infectious units doubling approximately every week.\footnote{We show in detail how further measurements in mouse lines expressing different levels of PrP along, with estimates of the average aggregate size, can be used to further dissect these rates and obtain mechanistic conclusions in a separate work.} 

The data of Alzheimer's disease are obtained from Gomez-Isla \textit{et al.}\cite{Gomez-Isla1997} and constitute post-mortem counts of the numbers of neurofibrillary tangles (NFT), the microscopically visible deposits of aggregated tau protein, in individuals with Alzheimer's disease. Since there is no controlled initiation event in these data as there is in the mouse data, the time axis denotes the number of years since onset of the symptoms. The choice of time zero is only required to combine the data points from different patients and does not affect our analysis as it can be compensated by $t_h$ in Eq.~\eqref{eq:logistic}, so does not affect the value of $\kappa$.
We find that $\kappa=0.2$ years$^{-1}$, two orders of magnitude slower than the rate for prion disease in mice extracted above\footnote{Again, a detailed analysis of several datasets on the accumulation of tau in Alzheimer's disease is provided in a separate work.}

\subsection*{Limitations and possible future extensions}
\textbf{Spatial inhomogeneities in aggregation}
Spatial inhomogeneities are normally not of relevance for \textit{in vitro} experiments, but they can be important in some living systems. If spatial variations need to be taken into account, meaning if spreading of aggregates throughout space is slow enough to limit the overall rate, then a reaction-diffusion type approach has to be employed instead\cite{Weickenmeier2018,Pandya2017,Raj2012} and the description we present here only accounts for the local reaction, not the time evolution of the whole system.

\noindent\textbf{Surfaces, co-factors and activated states}
In some systems, the presence of co-factors or activated states is believed to be important in enabling monomeric proteins to add to aggregates. There are two regimes one needs to consider here: (i) the activated state is present in steady state and co-factor is present at constant concentration in excess. In this case, conversion rates and co-factor concentrations can simply be subsumed into the existing rate constants, and the functional form of the system is not affected. The steady state assumption is generally a viable one, as aggregation in living system usually proceeds on a timescale much slower than monomeric protein turnover. Low concentrations of co-factor that limit rates could however be of importance, which leads us to the second regime. (ii) the rate is limited by a lack of co-factor or activating surface. Such a situation can be reflected in a decrease in the effective monomer dependence of the rates. It has been extensively discussed in Meisl et al.\cite{Meisl2017a} and Dear et al. \cite{Dear2020CatNature} and is amenable to the same approaches by simple replacing rate constants with effective rate constants. Thus, in both regimes, the approach we develop here should be sufficient to describe the kinetics.

\noindent\textbf{More complex models}
\textit{In vitro}, multiplication of aggregates normally only involves reactions of aggregates, possible with monomers, while \textit{in vivo} complex, indirect mechanisms have been proposed, for example aggregates induce an inflammatory response in a cell, which indirectly triggers aggregate formation in another cell. Our approach also works in the latter case, as long the assumptions of linearity in the $f(t,j)$ are fulfilled. For example, as long as the probability that one cell triggers aggregation elsewhere is proportional to some function of the form $\sum_j\alpha_jf(t,j)$, such as the total mass or total number of aggregates, it does not matter whether in reality this multiplication is induced via an inflammatory pathway.

\section*{Conclusions}
In this paper we have developed expressions, Eqs.~\eqref{eq:kappa_gen} and \eqref{eq:mu_gen}, that approximate well  the average aggregate size, the exponential rate of increase of aggregate quantities and its scaling with monomer concentration in a very general, broad class of multiplication-growth-type mechanisms. We moreover show that a logistic function is suitable for extracting the exponential rate from experimental data and thus infer the general class of mechanism that gave rise to the observed kinetics. The robust, general approach we present here is key in obtaining mechanistic insights and avoid over-fitting in the often noisy and sparse experimental data from complex living systems.

\newpage
\bibliographystyle{unsrt}

\bibliography{../../paper_database}

\newpage
\section*{Appendix}

\subsection*{Stationarity of the size distribution}

A key assumption in we make in order to solve Eqs.~\eqref{eq:dM_pre_ss} and \eqref{eq:dP_pre_ss} is that the normalized size distribution of biofilaments stays unchanged. This does not always hold, and in fact requires that a feedback term exists in the kinetics, whereby a larger species generates a smaller species. To see this we write the size distribution kinetic equations in matrix form:
\begin{align}
\dot{\bm{f}}(t)&=\bm{A}\bm{f}(t)+\bm{\alpha}\\
\bm{f}(t)&=\begin{pmatrix} f(t,2) \\ f(t,3) \\ \vdots \\ f(t,j) \\ \vdots \end{pmatrix},
\end{align}
where $\bm{\alpha}$ is a (constant) vector of constants reflecting the contribution of primary nucleation-like reactions and any external fluxes to the rates of formation of filaments of each length. Moreover, $\bm{A}$ is the matrix governing the linear relationships between the different filaments. The general solution to this nonhomogeneous ODE system, with zero initial condition, is given by:
\begin{equation}
\bm{f}(t)=\bm{\Psi}(t)\int_{0}^{t}\bm{\Psi}^{-1}(t')\bm{\alpha}dt',
\end{equation}
where $\bm{\Psi}$ is the matrix whose columns are the eigenvectors of $\bm{A}$. Now, where there are no feedback terms from e.g.\ secondary nucleation or fragmentation, $\bm{A}$ is lower triangular:
\begin{equation}
\bm{A}= \begin{pmatrix} -\mu_{1}-q_1 & 0 & 0 & 0 & \cdots \\ \mu_{1} & -\mu_{2}-q_2 & 0 & 0 & \cdots \\ a_{3,1} & \mu_{2} & -\mu_{3}-q_3 & 0 & \cdots \\ a_{4,1} & a_{4,2} & \mu_{3} & -\mu_{4}-q_4 & \ddots \\ \vdots & \vdots & \vdots & \ddots & \ddots \end{pmatrix},
\end{equation}
where $\mu_i$ is the rate of monomer addition to fibrils of size $i$, and $q_i$ accounts for clearance and dissociation. In such a matrix, the eigenvalues are simply the diagonal elements, all of which are negative here. For distinct $\mu_{i}$'s, each solution mode has the form:
\begin{equation}
z(t,j)=c_j\left(1-e^{-(\mu_j+q_j) t}\right).
\end{equation}
Thus the modes individually tend rapidly toward (positive) limiting values, and a steady state size distribution is never achieved. Note that since there are infinitely many $\mu$, many of the eigenvalues will be very similar in value, leading to very similar eigenvectors. The individual $f(t,j)$ are constructed as sums and differences of the first $j$ modes; they therefore take much longer to reach their limiting values with increasing size than do the individual modes.

This argument demonstrates that without feedback reaction steps to populate the upper triangular section of the matrix, a steady-state size distribution cannot be achieved. It does not, however, demonstrate that their presence is a sufficient condition. To do this, we must understand what such a distribution would result from. It essentially requires that one of the solution modes become much larger than the others at late enough time. This in turn requires that the largest eigenvalue is positive, since the resulting exponential growth of the associated mode will ensure its late-time dominance, no matter how small the difference between the two largest eigenvalues (exponential behaviour is guaranteed by linearity). Clearly, the timescale for attainment of steady-state is just the difference in magnitude of these two largest eigenvalues.

Unfortunately, the characteristic polynomial of a non-triangular matrix is not well-behaved as the dimension tends to infinity, due to sign changes. A different approach is thus likely necessary. We note, however, that stationarity has been proved for secondary nucleation and fragmentation with a size-independent elongation rate\cite{Michaels2015}, given a size-independent $k_+$.

\end{document}